\documentclass[a4paper,fleqn]{cas-dc}
\usepackage[numbers]{natbib}
\usepackage{booktabs}
\usepackage{adjustbox}
\usepackage{longtable}
\usepackage{caption}
\usepackage{mathtools}

\def\tsc#1{\csdef{#1}{\textsc{\lowercase{#1}}\xspace}}
\tsc{WGM}
\tsc{QE}
\tsc{EP}
\tsc{PMS}
\tsc{BEC}
\tsc{DE}
\begin{document}
\let\WriteBookmarks\relax
\def\floatpagepagefraction{1}
\def\textpagefraction{.001}
\let\printorcid\relax

\shorttitle{The role of intra- and inter-group Matthew effect in the social dilemma of public goods games}

\shortauthors{Chaoqian Wang {\it et~al.}}

\title [mode = title]{The role of intra- and inter-group Matthew effect in the social dilemma of public goods games}                      

\author[1]{Chaoqian Wang}
\ead{CqWang814921147@outlook.com}
\credit{xxx}
\cormark[1]
\cortext[cor1]{Corresponding author}

% Address/affiliation
\address[1]{Department of Computational and Data Sciences, George Mason University, Fairfax, VA 22030, USA}

% Here goes the abstract
\begin{abstract}
The Matthew effect describes the phenomenon where the rich tend to get richer. Such a success-driven mechanism has been studied in spatial public goods games in an inter-group way, where each individual's social power is enhanced across all groups. For instance, factors like knowledge can exert an advantage across various social contexts. In contrast, certain factors, especially local material goods, only enhance advantages within their current group. Building on this, we further explore the intra-group Matthew effect where the enhancement of social power is calculated separately in each group. Our findings indicate that the intra-group Matthew effect sustains cooperation more at high productivity, while the inter-group Matthew effect promotes cooperation at low productivity. Moreover, the mixture of the intra- and inter-group Matthew effect harms cooperation. This study provides insights into addressing social dilemmas by adjusting wealth accumulation across diverse social groups.
\end{abstract}

% Use if graphical abstract is present
% \begin{graphicalabstract}
% \includegraphics{figs/grabs.pdf}
% \end{graphicalabstract}

% Research highlights
% \begin{highlights}
% \item The impact of inertia is studied in public goods games in the weak selection limit
% \item Analytical results are derived for the critical synergy factor of cooperation success
% \item Inertia links the cooperation success conditions for different updating rules
% \item Theoretical results are confirmed by Monte Carlo simulations
% \end{highlights}


\maketitle
\section{Introduction}\label{sec_intro}
Matthew 25:29 (New Revised Standard Version) states, ``For to all those who have, more will be given, and they will have an abundance; but from those who have nothing, even what they have will be taken away.'' In the real world, it is evident that wealthy individuals have greater opportunities to become richer, while the poor have limited prospects of improving their financial situations. This phenomenon is known as the Matthew effect~\cite{merton1968matthew,perc2014matthew}. The role of the Matthew effect has been studied in the social dilemma of evolutionary public goods games~\cite{perc2011success}.

Social dilemmas refer to the conflict between individual and collective interests~\cite{sigmund2010calculus}. While cooperation (also known as prosocial behavior) benefits everyone in a group, individual defection can benefit oneself at the expense of others, resulting in a reduction in average benefits for all group members, leading to the tragedy of the commons~\cite{hardin1968tragedy}. However, cooperation can thrive in human society~\cite{nowak2006evolutionary}, and several explanations have been proposed for this phenomenon~\cite{nowak2006five}, including network reciprocity, which has been studied by researchers from evolutionary biology to statistical physics~\cite{perc2017statistical} and from computer simulations~\cite{nowak1992evolutionary,nowak1993spatial,szabo1998evolutionary} to theoretical analyses~\cite{ohtsuki2006simple,lieberman2005evolutionary,allen2017evolutionary}. 

It is a natural idea to consider the role of the Matthew effect in multiplayer games instead of two-player games because the former involves the allocation of public goods and the enhancement of wealth. The emergence of cooperation in multiplayer games has become increasingly popular over the past few decades~\cite{chen2017evolutionary,luo2021evolutionary,wang2022involution}, and group interactions~\cite{perc2013evolutionary} play a significant role in higher-order population structures~\cite{burgio2020evolution,alvarez2021evolutionary,pan2023heterogeneous,duh2023unexpected}. The fundamental model type is the public goods game, in which cooperators contribute to the group, producing more than the cost, but defectors can share the fruits equally~\cite{szabo2002phase}. In other words, in a classic public goods game group, everyone shares an equal amount of fruits. However, introducing the Matthew effect mentioned earlier, those receiving more can share more. It was found that this can have an intriguing effect on the cooperation level: the Matthew effect in distributing fruits promotes cooperation but also preserves defection~\cite{perc2011success,hu2023cumulative}. Similar to this success-driven mechanism in distribution, other success-driven mechanisms in different aspects of evolutionary dynamics have been studied, such as success-driven migration~\cite{helbing2009outbreak,jiang2010role,liu2012does,buesser2013opportunistic}, success-driven participation~\cite{chen2016individual}, success-driven group formation~\cite{szolnoki2016cooperation}, and success-driven multigames~\cite{szolnoki2014coevolutionary}. Opposite assumptions from the Matthew effect have also been investigated, such as the situation where rich people contribute more~\cite{tu2018contribution}.

A widely accepted notion is that inter-group reciprocity, where an individual participates in multiple groups and engages in games in each, plays a crucial role in spatial public goods games~\cite{nowak2010evolutionary,su2018understanding, su2019spatial}. When coupling groups, apparently similar things can differ~\cite{wang2022reversed}. Previous research has investigated various inter-group mechanisms in public goods games, such as group competition~\cite{cheikbossian2012collective} and inter-group selection~\cite{eckel2016group, wang2021inter}. Some studies have also examined scenarios where individuals use different strategies with different neighbors~\cite{su2016interactive, su2017evolutionary}.

Similarly, as an additional mechanism, the Matthew effect can also be considered separately in different groups. On the one hand, an individual can accumulate social power in different groups independently. On the other hand, an individual can also possess a unified social power through all its groups, as previous research has shown~\cite{perc2011success}. More generally, the intra- and inter-group social power can be blended, and the impact of their proportions on the role they play can be explored. The idea of differentiating the intra- and inter-group Matthew effects captures a detail of the real world: people are involved in multiple social groups and interact with members in each. For example, a student attends different classes at a university, and their advantages accrue separately or jointly in various classes, depending on the relevance of the knowledge. Similarly, a multinational corporation operates in different countries, and its advantages in each country may accumulate separately or jointly, depending on the trade policies.

In this study, we establish a model to describe the intra- and inter-group Matthew effects in the spatial public goods game and investigate their roles separately and collectively. It should be noted that we use synchronous updates~\cite{nowak1992evolutionary,nowak1993spatial,wang2021public,wang2022between} in this work. Although most previous studies adopt asynchronous updates, the synchronous approach is more intuitive in describing specific mechanisms~\cite{liu2010memory,dong2019memory}. The results do not differ significantly in quality~\cite{shu2018memory}.

\section{Model}\label{secmodel}
The spatial public goods game is conducted on an $L\times L$ square lattice, where each node is occupied by an agent. The agent set is denoted by $\mathcal{N}$, and its size is the population $N$, which yields $|\mathcal{N}|\equiv N=L^2$. Each agent $i\in\mathcal{N}$ is the center of a group $\Omega_i$, which contains the centering agent $i$ and its four nearest neighbors $j\in\Omega_i\setminus \{i\}$. In this work, we set the group size to $G=5$. Each agent $i$ is involved in $G$ groups, centered on itself and its neighbors, $\Omega_j$ for $j\in\Omega_i$.

The payoff received by agent $i$ from the group centered on $j$ at time $t$ is denoted by $\pi_{i,t}^j$, calculated as follows:
\begin{equation}
\pi_{i,t}^j=M_{i,j,t}r\sum_{k\in\Omega_j}s_{k,t}c-s_{i,t}c,
\end{equation}
where $s_{i,t}$ denotes the strategy of agent $i$ at time $t$. If $s_{i,t}=1$, agent $i$ cooperates by contributing $c$ ($c>0$) to the group $j$, while if $s_{i,t}=0$, agent $i$ defects and makes no contribution. The contributions to group $j$ by players $k\in\Omega_j$ are enlarged by the productivity factor $r$ ($r>1$) and allocated according to the Matthew effect $M_{i,j,t}$, which we will define later.

The average payoff of agent $i$ received from its $G$ groups at time $t$ is denoted by $\pi_{i,t}$, yielding
\begin{equation}
\pi_{i,t}=\frac{1}{G}\sum_{j\in\Omega_i}\pi_{i,t}^j.
\end{equation}

At each time step $t$, the agents' strategies $s_{i,t}$ are updated synchronously. For each agent $i$, one of its neighbors $j\in\Omega_i\setminus \{i\}$ is selected randomly, and agent $i$ adopts $j$'s strategy $s_{j,t}$ with probability
\begin{equation}
W(s_{i,t+1}\gets s_{j,t})=\frac{1}{1+\mathrm{e}^{-(\pi_{j,t}-\pi_{i,t})/\kappa}},
\end{equation}
where $\kappa$ is the noise parameter, set to $\kappa=0.1$ in this work. If the update is not successful, agent $i$ keeps its current strategy, $s_{i,t+1}\gets s_{i,t}$.

To define the Matthew effect, we introduce the concept of social power, which is determined by an agent's payoff, both within and between groups. First, an agent's intra-group social power varies across groups. Specifically, agent $i$'s intra-group power in group $j$ at time $t+1$ is based on its payoff from group $j$ at time $t$, as given by Eq.~(\ref{eq_pintra}):
\begin{equation}\label{eq_pintra}
P_{i,j,t+1}^{(\text{intra})}=\pi_{i,t}^j.
\end{equation}
Second, an agent's inter-group social power is the average of its intra-group power across all of its $G$ groups. In other words, it is simply the agent's payoff at time $t$, as expressed by Eq.~(\ref{eq_pinter}):
\begin{equation}\label{eq_pinter}
P_{i,t+1}^{(\text{inter})}=\frac{1}{G}\sum_{j\in\Omega_i}P_{i,j,t+1}^{(\text{intra})}=\pi_{i,t}.
\end{equation}
The social power of agent $i$ is a weighted combination of its intra- and inter-group power, as given by Eq.~(\ref{eq_p}):
\begin{equation}\label{eq_p}
P_{i,j,t+1}=aP_{i,j,t+1}^{(\text{intra})}+(1-a)P_{i,t+1}^{(\text{inter})},
\end{equation}
where $0\leq a \leq 1$ determines the relative weight of intra-group power versus inter-group power.

The Matthew effect, which determines the allocation of public goods to agent $i$ in group $j$, is proportional to the agent's social power. Specifically, the higher an agent's social power, the more public goods are allocated to it. This is captured by Eq.~(\ref{eq_matthew}):
\begin{equation}\label{eq_matthew}
M_{i,j,t+1}=\frac{\mathrm{e}^{wP_{i,j,t+1}}}{\sum_{k\in\Omega_j}\mathrm{e}^{wP_{k,j,t+1}}},
\end{equation}
where $w\geq 0$ determines the strength of the Matthew effect. A larger value of $w$ corresponds to a stronger Matthew effect, in which agents with higher payoffs receive a larger share of the public good. When $w=0$, Eq.~(\ref{eq_matthew}) reduces to $M_{i,j,t}\equiv 1/G$ for all $i\in\mathcal{N}$, $j\in\Omega_i$, and $t\in\mathbf{N}^*$, which corresponds to the classic public goods game.

In our model, the Matthew effect is self-cumulative, as described by Eqs.~(\ref{eq_pintra})--(\ref{eq_pinter}). Specifically, the social power at time $t$ affects the allocation of public goods at time $t+1$, which in turn affects the social power at time $t+1$ and the allocation at time $t+2$, and so on. This automatic iteration process obviates the need for storing data from every previous step.

\section{Results and discussion}
The fixed secondary parameters of this study are population size $N=100\times 100=10^4$, group size $G=5$, and selection noise $\kappa=0.1$. The main parameters are productivity $r$, the strength of the Matthew effect $w$, and the proportion of the intra-group Matthew effects $a$, whose values are flexible. 

At the start of the simulation, each agent $i \in \mathcal{N}$ is assigned a random strategy of cooperation ($s_{i,0}=1$) or defection ($s_{i,0}=0$), with the proportion of cooperators being approximately $1/2$. The social power $P_{i,j,0}$ for each agent in their respective groups $j \in \Omega_i$ is initialized to 0, resulting in $M_{i,j,0}=1/G$ and an equal distribution of public goods at the first time step.

The system is then allowed to evolve from $t=0$ to $t=10^5$, and the proportion of cooperators $\sum_{i\in\mathcal{N}}s_{i,t}/N$ is calculated at each time step $t$. The average proportion of cooperators in the last $2\times 10^4$ time steps is used to define the cooperation level ($\rho_C$) under the assigned parameter values, $\rho_C=\sum_{t=8\times 10^4+1}^{10^5}\sum_{i\in\mathcal{N}}s_{i,t}/N/(2\times 10^4)$.

\subsection{The role of the inter-group Matthew effect}
\begin{figure*}
\centering
\includegraphics[width=.6\textwidth]{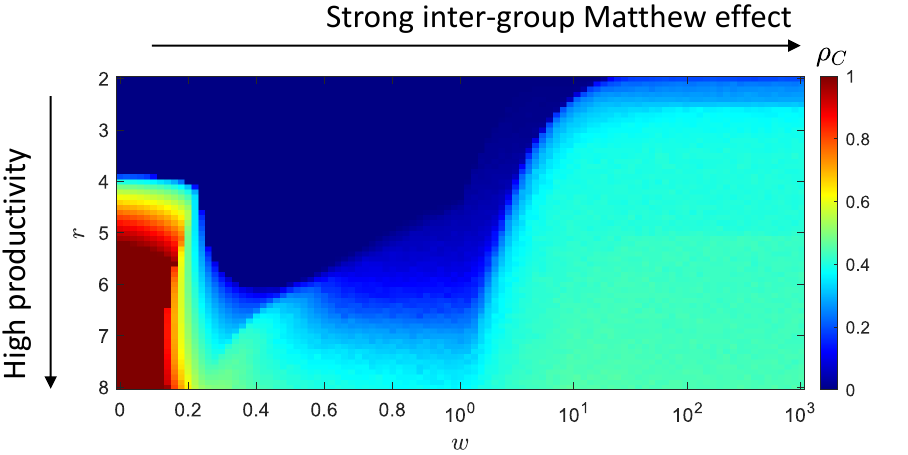}
\caption{Cooperation level $\rho_C$ as a function of productivity $r$ and the strength of inter-group Matthew effect $w$, with fixed intra-group proportion $a=0$. The $w$-axis scale is linear for $0\leq w\leq 1$ and logarithmic for $10^0\leq w \leq 10^3$. The results are qualitatively similar to the ones in \cite{perc2011success}: a strong inter-group Matthew effect promotes the emergence of cooperation when productivity is low, but also preserves defection when productivity is high. Meanwhile, a moderate inter-group Matthew effect disfavors cooperation most.}
\label{fig_rwa0}
\end{figure*}

We begin by examining a special case where only the inter-group Matthew effect is present. When $a=0$, the level of cooperation, $\rho_C$, is plotted as a function of productivity $r$ and the strength of the Matthew effect $w$ in Fig.~\ref{fig_rwa0}. To facilitate a comprehensive presentation of the results, we have rescaled the $w$-axis logarithmically for $w>1$, while the scale in $0<w<1$ is linear. When $w=0$, the results coincide with those of the classic synchronous spatial public goods game. However, when $w>0$, we can observe the impact of the inter-group Matthew effect's strength by increasing $w$.

As illustrated in Fig.~\ref{fig_rwa0}, the inter-group Matthew effect promotes cooperation at low productivity ($r\lesssim 4$), with cooperation emerging at $w\gtrsim 1$. However, at intermediate or high productivity ($r\gtrsim 4$), the inter-group Matthew effect inhibits cooperation. Interestingly, the inhibition of cooperation is most significant when the inter-group Matthew effect is moderate. As the inter-group Matthew effect becomes stronger, the cooperation level rebounds, albeit remaining lower than in the absence of the Matthew effect. Notably, for intermediate productivity ($4\lesssim r\lesssim 6$), cooperation disappears entirely when the effect strength is moderate. Our findings are consistent with previous work~\cite{perc2011success} that investigated the pure inter-group Matthew effect and concluded that it promotes cooperation at low productivity but sustains defection at high productivity, albeit for a different implementation of our approach in modeling.

\subsection{The role of the intra-group Matthew effect}
\begin{figure*}
\centering
\includegraphics[width=.6\textwidth]{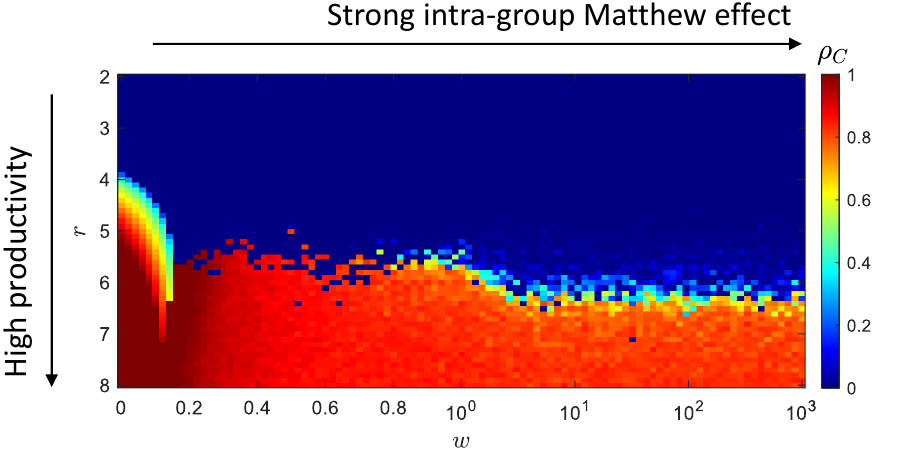}
\caption{Cooperation level $\rho_C$ as a function of productivity $r$ and the strength of intra-group Matthew effect $w$, with fixed intra-group proportion $a=1$. The $w$-axis scale is linear for $0\leq w\leq 1$ and logarithmic for $10^0\leq w \leq 10^3$. A strong intra-group Matthew effect impedes cooperation, and, compared to this, a moderate intra-group Matthew effect can sustain cooperation more.}
\label{fig_rwa1}
\end{figure*}

Next, we analyze the impact of the intra-group Matthew effect by setting $a=1$ and examine the relationship between the intra-group effect's strength and the cooperation level. Fig.~\ref{fig_rwa1} illustrates the cooperation level as a function of productivity and the strength of the intra-group Matthew effect, with the $w$-axis scaled similarly to Fig.~\ref{fig_rwa0}. When the intra-group Matthew effect is present ($w>0$), we can observe its role by increasing $w$.

In contrast to the inter-group Matthew effect, a certain level of intra-group Matthew effect ($w\gtrsim 0.2$) can only promote cooperation at high productivity ($r\gtrsim 5$). At moderate strength ($0.2\lesssim w\lesssim 1$), neither the intra-group nor the inter-group Matthew effect can promote cooperation at low productivity, but the intra-group Matthew effect can sustain higher levels of cooperation at high productivity. At higher strength ($w\gtrsim 10^0$), the intra-group Matthew effect cannot promote cooperation at low productivity as effectively as the inter-group effect, but it can maintain higher levels of cooperation at high productivity. In conclusion, while the intra-group Matthew effect cannot promote cooperation at low productivity, it can retain more cooperation at high productivity compared to the inter-group Matthew effect. This advantage is most pronounced at a moderate strength of the Matthew effect.

To explain how the intra-group Matthew effect maintains cooperation better than the inter-group effect at high productivity, we can imagine the interface between cooperation and defection clusters and consider the role of the intra-group Matthew effect. In a public goods game within a single group, defectors receive higher payoffs than cooperators because of the extra cost $c$ incurred by cooperators. Consequently, when allocation in the next moment within a group depends solely on the current payoff, individuals who adopt the defective strategy are always allocated a larger proportion $M_{i,j,t+1}$ of the public good than cooperative individuals. Under the high productivity of the classic model, cooperative clusters expand rapidly. Defectors who can allocate more public goods in the next moment mostly become cooperators in the next moment. As a result, they are allocated the same proportion of public goods in the group facing the defection region and more public goods in the group facing the cooperation region due to the effect of defection in the previous moment. This transformation of defectors into cooperators encourages adjacent defectors to become more willing to transform into cooperators, which preserves the rapid expansion of cooperative clusters.

\subsection{The combination of the intra- and inter-group Matthew effect}
\begin{figure*}
\centering
\includegraphics[width=\textwidth]{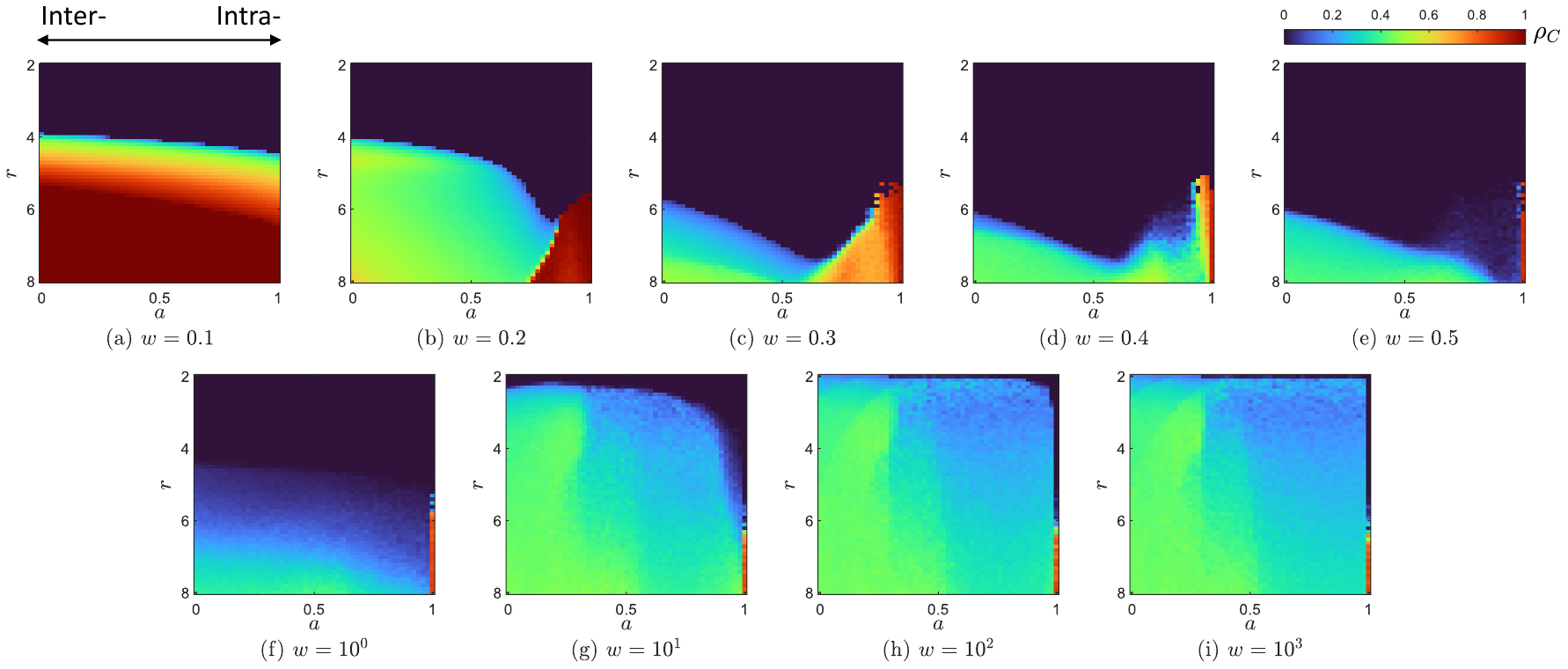}
\caption{Cooperation level $\rho_C$ as a function of productivity $r$ and the intra-group proportions $a$ at different strength of intra-group Matthew effect $w$. The values of $w$ are presented in the subtitles. Only the inter-group Matthew effect plays a role when $a=0$. Only the intra-group Matthew effect plays a role when $a=1$. The intra-group and inter-group Matthew effects are mixed when $0<a<1$. Generally speaking, cooperation is disfavored when the intra- and inter-group Matthew effects are both present.}
\label{fig_ra}
\end{figure*}

Finally, we investigate the combined effects of the intra- and inter-group Matthew effects, where the intra-group social power is represented as a proportion of $a$ and the inter-group social power as a proportion of $1-a$. We use Fig.~\ref{fig_ra} to display the cooperation level as a function of productivity $r$ and the proportion of the intra-group Matthew effect $a$ at different strengths of the Matthew effect $w$.

When the strength of the Matthew effect is moderate [Fig.~\ref{fig_ra}(b), (c), and (d)], we find that the intra-group Matthew effect has a dominant range where the cooperation level is high, particularly in the region where $a$ is large. The cooperation level varies continuously within this region. Similarly, the inter-group Matthew effect has its dominant range of smaller $a$. The cooperation level varies sharply on the adjoining interface of the dominant range of both effects, which is the most prominent in Fig.~\ref{fig_ra}(b).

In contrast, when the strength of the Matthew effect is large, we observe that the intra-group Matthew effect only manifests when it exists alone ($a=1$). In other cases where $a<1$, the change in cooperation level is dominated by the inter-group Matthew effect, which varies continuously at low cooperation levels. Overall, in each panel of Fig.~\ref{fig_ra}, the cooperation level is high at the ends of the $a$-axis and low in the middle. These results suggest that a combination of intra- and inter-group Matthew effects is detrimental to cooperation.
 
\section{Conclusion}
In conclusion, this study expands upon the concept introduced in~\cite{perc2011success} and examines the impact of the intra- and inter-group Matthew effect separately and in combination. Our findings indicate that the inter-group Matthew effect inhibits cooperation at a moderate level but can sustain it to some degree when it is strong. In comparison, the intra-group Matthew effect encourages cooperation, particularly when productivity is high, due to the rapid spread of cooperative strategies at cluster boundaries. However, the combination of both intra- and inter-group Matthew effects has a negative impact on cooperation.

The numerical simulations in this study have the potential to provide insights for policy analysis on wealth distribution. As previously noted, individuals participate in different games within various social relationships, and the Matthew effect is a factor that cannot be ignored in the real world. To address this issue, we suggest adjusting the role of the wealth enhancement effect based on the current level of social productivity. At low productivity, the Matthew effect should be transferable between different social relationships, and success in one group should facilitate success in all groups. At high productivity, the Matthew effect should not be correlated across different social relationships, and success in one group should only increase the likelihood of success in the same group. In any case, we recommend minimizing the mixing of these two approaches.

\section*{Declaration of competing interest}
None.

\section*{Code availability}
The Matlab code for the numerical simulations is available upon request from the corresponding author.

% \bibliographystyle{elsarticle-num-names}
% \bibliography{cas-refs}

\begin{thebibliography}{46}
\expandafter\ifx\csname natexlab\endcsname\relax\def\natexlab#1{#1}\fi
\providecommand{\url}[1]{\texttt{#1}}
\providecommand{\href}[2]{#2}
\providecommand{\path}[1]{#1}
\providecommand{\DOIprefix}{doi:}
\providecommand{\ArXivprefix}{arXiv:}
\providecommand{\URLprefix}{URL: }
\providecommand{\Pubmedprefix}{pmid:}
\providecommand{\doi}[1]{\href{http://dx.doi.org/#1}{\path{#1}}}
\providecommand{\Pubmed}[1]{\href{pmid:#1}{\path{#1}}}
\providecommand{\bibinfo}[2]{#2}
\ifx\xfnm\relax \def\xfnm[#1]{\unskip,\space#1}\fi
%Type = Article
\bibitem[{Merton(1968)}]{merton1968matthew}
\bibinfo{author}{R.~K. Merton},
\newblock \bibinfo{title}{The {Matthew} effect in science: The reward and
  communication systems of science are considered.},
\newblock \bibinfo{journal}{Science} \bibinfo{volume}{159}
  (\bibinfo{year}{1968}) \bibinfo{pages}{56--63}.
%Type = Article
\bibitem[{Perc(2014)}]{perc2014matthew}
\bibinfo{author}{M.~Perc},
\newblock \bibinfo{title}{The {Matthew effect} in empirical data},
\newblock \bibinfo{journal}{Journal of The Royal Society Interface}
  \bibinfo{volume}{11} (\bibinfo{year}{2014}) \bibinfo{pages}{20140378}.
%Type = Article
\bibitem[{Perc(2011)}]{perc2011success}
\bibinfo{author}{M.~Perc},
\newblock \bibinfo{title}{Success-driven distribution of public goods promotes
  cooperation but preserves defection},
\newblock \bibinfo{journal}{Physical Review E} \bibinfo{volume}{84}
  (\bibinfo{year}{2011}) \bibinfo{pages}{037102}.
%Type = Book
\bibitem[{Sigmund(2010)}]{sigmund2010calculus}
\bibinfo{author}{K.~Sigmund}, \bibinfo{title}{The calculus of selfishness},
  \bibinfo{publisher}{Princeton University Press}, \bibinfo{year}{2010}.
%Type = Article
\bibitem[{Hardin(1968)}]{hardin1968tragedy}
\bibinfo{author}{G.~Hardin},
\newblock \bibinfo{title}{The tragedy of the commons},
\newblock \bibinfo{journal}{Science} \bibinfo{volume}{162}
  (\bibinfo{year}{1968}) \bibinfo{pages}{1243--1248}.
%Type = Book
\bibitem[{Nowak(2006{\natexlab{a}})}]{nowak2006evolutionary}
\bibinfo{author}{M.~A. Nowak}, \bibinfo{title}{Evolutionary dynamics: exploring
  the equations of life}, \bibinfo{publisher}{Harvard University Press},
  \bibinfo{year}{2006}{\natexlab{a}}.
%Type = Article
\bibitem[{Nowak(2006{\natexlab{b}})}]{nowak2006five}
\bibinfo{author}{M.~A. Nowak},
\newblock \bibinfo{title}{Five rules for the evolution of cooperation},
\newblock \bibinfo{journal}{Science} \bibinfo{volume}{314}
  (\bibinfo{year}{2006}{\natexlab{b}}) \bibinfo{pages}{1560--1563}.
%Type = Article
\bibitem[{Perc et~al.(2017)Perc, Jordan, Rand, Wang, Boccaletti, and
  Szolnoki}]{perc2017statistical}
\bibinfo{author}{M.~Perc}, \bibinfo{author}{J.~J. Jordan},
  \bibinfo{author}{D.~G. Rand}, \bibinfo{author}{Z.~Wang},
  \bibinfo{author}{S.~Boccaletti}, \bibinfo{author}{A.~Szolnoki},
\newblock \bibinfo{title}{Statistical physics of human cooperation},
\newblock \bibinfo{journal}{Physics Reports} \bibinfo{volume}{687}
  (\bibinfo{year}{2017}) \bibinfo{pages}{1--51}.
%Type = Article
\bibitem[{Nowak and May(1992)}]{nowak1992evolutionary}
\bibinfo{author}{M.~A. Nowak}, \bibinfo{author}{R.~M. May},
\newblock \bibinfo{title}{Evolutionary games and spatial chaos},
\newblock \bibinfo{journal}{Nature} \bibinfo{volume}{359}
  (\bibinfo{year}{1992}) \bibinfo{pages}{826--829}.
%Type = Article
\bibitem[{Nowak and May(1993)}]{nowak1993spatial}
\bibinfo{author}{M.~A. Nowak}, \bibinfo{author}{R.~M. May},
\newblock \bibinfo{title}{The spatial dilemmas of evolution},
\newblock \bibinfo{journal}{International Journal of Bifurcation and Chaos}
  \bibinfo{volume}{3} (\bibinfo{year}{1993}) \bibinfo{pages}{35--78}.
%Type = Article
\bibitem[{Szab{\'o} and T{\H{o}}ke(1998)}]{szabo1998evolutionary}
\bibinfo{author}{G.~Szab{\'o}}, \bibinfo{author}{C.~T{\H{o}}ke},
\newblock \bibinfo{title}{Evolutionary prisoner’s dilemma game on a square
  lattice},
\newblock \bibinfo{journal}{Physical Review E} \bibinfo{volume}{58}
  (\bibinfo{year}{1998}) \bibinfo{pages}{69}.
%Type = Article
\bibitem[{Ohtsuki et~al.(2006)Ohtsuki, Hauert, Lieberman, and
  Nowak}]{ohtsuki2006simple}
\bibinfo{author}{H.~Ohtsuki}, \bibinfo{author}{C.~Hauert},
  \bibinfo{author}{E.~Lieberman}, \bibinfo{author}{M.~A. Nowak},
\newblock \bibinfo{title}{A simple rule for the evolution of cooperation on
  graphs and social networks},
\newblock \bibinfo{journal}{Nature} \bibinfo{volume}{441}
  (\bibinfo{year}{2006}) \bibinfo{pages}{502--505}.
%Type = Article
\bibitem[{Lieberman et~al.(2005)Lieberman, Hauert, and
  Nowak}]{lieberman2005evolutionary}
\bibinfo{author}{E.~Lieberman}, \bibinfo{author}{C.~Hauert},
  \bibinfo{author}{M.~A. Nowak},
\newblock \bibinfo{title}{Evolutionary dynamics on graphs},
\newblock \bibinfo{journal}{Nature} \bibinfo{volume}{433}
  (\bibinfo{year}{2005}) \bibinfo{pages}{312--316}.
%Type = Article
\bibitem[{Allen et~al.(2017)Allen, Lippner, Chen, Fotouhi, Momeni, Yau, and
  Nowak}]{allen2017evolutionary}
\bibinfo{author}{B.~Allen}, \bibinfo{author}{G.~Lippner},
  \bibinfo{author}{Y.-T. Chen}, \bibinfo{author}{B.~Fotouhi},
  \bibinfo{author}{N.~Momeni}, \bibinfo{author}{S.-T. Yau},
  \bibinfo{author}{M.~A. Nowak},
\newblock \bibinfo{title}{Evolutionary dynamics on any population structure},
\newblock \bibinfo{journal}{Nature} \bibinfo{volume}{544}
  (\bibinfo{year}{2017}) \bibinfo{pages}{227--230}.
%Type = Article
\bibitem[{Chen et~al.(2017)Chen, Gracia-L{\'a}zaro, Li, Wang, and
  Moreno}]{chen2017evolutionary}
\bibinfo{author}{W.~Chen}, \bibinfo{author}{C.~Gracia-L{\'a}zaro},
  \bibinfo{author}{Z.~Li}, \bibinfo{author}{L.~Wang},
  \bibinfo{author}{Y.~Moreno},
\newblock \bibinfo{title}{Evolutionary dynamics of $n$-person {Hawk-Dove}
  games},
\newblock \bibinfo{journal}{Scientific Reports} \bibinfo{volume}{7}
  (\bibinfo{year}{2017}) \bibinfo{pages}{1--10}.
%Type = Article
\bibitem[{Luo et~al.(2021)Luo, Liu, and Chen}]{luo2021evolutionary}
\bibinfo{author}{Q.~Luo}, \bibinfo{author}{L.~Liu}, \bibinfo{author}{X.~Chen},
\newblock \bibinfo{title}{Evolutionary dynamics of cooperation in the
  $n$-person stag hunt game},
\newblock \bibinfo{journal}{Physica D: Nonlinear Phenomena}
  \bibinfo{volume}{424} (\bibinfo{year}{2021}) \bibinfo{pages}{132943}.
%Type = Article
\bibitem[{Wang and Szolnoki(2022)}]{wang2022involution}
\bibinfo{author}{C.~Wang}, \bibinfo{author}{A.~Szolnoki},
\newblock \bibinfo{title}{Involution game with spatio-temporal heterogeneity of
  social resources},
\newblock \bibinfo{journal}{Applied Mathematics and Computation}
  \bibinfo{volume}{430} (\bibinfo{year}{2022}) \bibinfo{pages}{127307}.
%Type = Article
\bibitem[{Perc et~al.(2013)Perc, G{\'o}mez-Gardenes, Szolnoki, Flor{\'\i}a, and
  Moreno}]{perc2013evolutionary}
\bibinfo{author}{M.~Perc}, \bibinfo{author}{J.~G{\'o}mez-Gardenes},
  \bibinfo{author}{A.~Szolnoki}, \bibinfo{author}{L.~M. Flor{\'\i}a},
  \bibinfo{author}{Y.~Moreno},
\newblock \bibinfo{title}{Evolutionary dynamics of group interactions on
  structured populations: a review},
\newblock \bibinfo{journal}{Journal of the Royal Society Interface}
  \bibinfo{volume}{10} (\bibinfo{year}{2013}) \bibinfo{pages}{20120997}.
%Type = Article
\bibitem[{Burgio et~al.(2020)Burgio, Matamalas, G{\'o}mez, and
  Arenas}]{burgio2020evolution}
\bibinfo{author}{G.~Burgio}, \bibinfo{author}{J.~T. Matamalas},
  \bibinfo{author}{S.~G{\'o}mez}, \bibinfo{author}{A.~Arenas},
\newblock \bibinfo{title}{Evolution of cooperation in the presence of
  higher-order interactions: From networks to hypergraphs},
\newblock \bibinfo{journal}{Entropy} \bibinfo{volume}{22}
  (\bibinfo{year}{2020}) \bibinfo{pages}{744}.
%Type = Article
\bibitem[{Alvarez-Rodriguez et~al.(2021)Alvarez-Rodriguez, Battiston,
  de~Arruda, Moreno, Perc, and Latora}]{alvarez2021evolutionary}
\bibinfo{author}{U.~Alvarez-Rodriguez}, \bibinfo{author}{F.~Battiston},
  \bibinfo{author}{G.~F. de~Arruda}, \bibinfo{author}{Y.~Moreno},
  \bibinfo{author}{M.~Perc}, \bibinfo{author}{V.~Latora},
\newblock \bibinfo{title}{Evolutionary dynamics of higher-order interactions in
  social networks},
\newblock \bibinfo{journal}{Nature Human Behaviour} \bibinfo{volume}{5}
  (\bibinfo{year}{2021}) \bibinfo{pages}{586--595}.
%Type = Article
\bibitem[{Pan et~al.(2023)Pan, Zhang, Han, and Huang}]{pan2023heterogeneous}
\bibinfo{author}{J.~Pan}, \bibinfo{author}{L.~Zhang}, \bibinfo{author}{W.~Han},
  \bibinfo{author}{C.~Huang},
\newblock \bibinfo{title}{Heterogeneous investment promotes cooperation in
  spatial public goods game on hypergraphs},
\newblock \bibinfo{journal}{Physica A: Statistical Mechanics and its
  Applications} \bibinfo{volume}{609} (\bibinfo{year}{2023})
  \bibinfo{pages}{128400}.
%Type = Article
\bibitem[{Duh et~al.(2023)Duh, Gosak, and Perc}]{duh2023unexpected}
\bibinfo{author}{M.~Duh}, \bibinfo{author}{M.~Gosak},
  \bibinfo{author}{M.~Perc},
\newblock \bibinfo{title}{Unexpected paths to cooperation on tied hyperbolic
  networks},
\newblock \bibinfo{journal}{Europhysics Letters} \bibinfo{volume}{142}
  (\bibinfo{year}{2023}) \bibinfo{pages}{62002}.
%Type = Article
\bibitem[{Szab{\'o} and Hauert(2002)}]{szabo2002phase}
\bibinfo{author}{G.~Szab{\'o}}, \bibinfo{author}{C.~Hauert},
\newblock \bibinfo{title}{Phase transitions and volunteering in spatial public
  goods games},
\newblock \bibinfo{journal}{Physical Review Letters} \bibinfo{volume}{89}
  (\bibinfo{year}{2002}) \bibinfo{pages}{118101}.
%Type = Article
\bibitem[{Hu et~al.(2023)Hu, Shi, Tao, and Perc}]{hu2023cumulative}
\bibinfo{author}{K.~Hu}, \bibinfo{author}{L.~Shi}, \bibinfo{author}{Y.~Tao},
  \bibinfo{author}{M.~Perc},
\newblock \bibinfo{title}{Cumulative advantage is a double-edge sword for
  cooperation},
\newblock \bibinfo{journal}{Europhysics Letters} \bibinfo{volume}{142}
  (\bibinfo{year}{2023}) \bibinfo{pages}{21001}.
%Type = Article
\bibitem[{Helbing and Yu(2009)}]{helbing2009outbreak}
\bibinfo{author}{D.~Helbing}, \bibinfo{author}{W.~Yu},
\newblock \bibinfo{title}{The outbreak of cooperation among success-driven
  individuals under noisy conditions},
\newblock \bibinfo{journal}{Proceedings of the National Academy of Sciences}
  \bibinfo{volume}{106} (\bibinfo{year}{2009}) \bibinfo{pages}{3680--3685}.
%Type = Article
\bibitem[{Jiang et~al.(2010)Jiang, Wang, Lai, and Wang}]{jiang2010role}
\bibinfo{author}{L.-L. Jiang}, \bibinfo{author}{W.-X. Wang},
  \bibinfo{author}{Y.-C. Lai}, \bibinfo{author}{B.-H. Wang},
\newblock \bibinfo{title}{Role of adaptive migration in promoting cooperation
  in spatial games},
\newblock \bibinfo{journal}{Physical Review E} \bibinfo{volume}{81}
  (\bibinfo{year}{2010}) \bibinfo{pages}{036108}.
%Type = Article
\bibitem[{Liu et~al.(2012)Liu, Chen, Zhang, Tao, and Wang}]{liu2012does}
\bibinfo{author}{Y.~Liu}, \bibinfo{author}{X.~Chen},
  \bibinfo{author}{L.~Zhang}, \bibinfo{author}{F.~Tao},
  \bibinfo{author}{L.~Wang},
\newblock \bibinfo{title}{Does migration cost influence cooperation among
  success-driven individuals?},
\newblock \bibinfo{journal}{Chaos, Solitons \& Fractals} \bibinfo{volume}{45}
  (\bibinfo{year}{2012}) \bibinfo{pages}{1301--1308}.
%Type = Article
\bibitem[{Buesser et~al.(2013)Buesser, Tomassini, and
  Antonioni}]{buesser2013opportunistic}
\bibinfo{author}{P.~Buesser}, \bibinfo{author}{M.~Tomassini},
  \bibinfo{author}{A.~Antonioni},
\newblock \bibinfo{title}{Opportunistic migration in spatial evolutionary
  games},
\newblock \bibinfo{journal}{Physical Review E} \bibinfo{volume}{88}
  (\bibinfo{year}{2013}) \bibinfo{pages}{042806}.
%Type = Article
\bibitem[{Chen and Szolnoki(2016)}]{chen2016individual}
\bibinfo{author}{X.~Chen}, \bibinfo{author}{A.~Szolnoki},
\newblock \bibinfo{title}{Individual wealth-based selection supports
  cooperation in spatial public goods games},
\newblock \bibinfo{journal}{Scientific Reports} \bibinfo{volume}{6}
  (\bibinfo{year}{2016}) \bibinfo{pages}{1--8}.
%Type = Article
\bibitem[{Szolnoki and Chen(2016)}]{szolnoki2016cooperation}
\bibinfo{author}{A.~Szolnoki}, \bibinfo{author}{X.~Chen},
\newblock \bibinfo{title}{Cooperation driven by success-driven group
  formation},
\newblock \bibinfo{journal}{Physical Review E} \bibinfo{volume}{94}
  (\bibinfo{year}{2016}) \bibinfo{pages}{042311}.
%Type = Article
\bibitem[{Szolnoki and Perc(2014)}]{szolnoki2014coevolutionary}
\bibinfo{author}{A.~Szolnoki}, \bibinfo{author}{M.~Perc},
\newblock \bibinfo{title}{Coevolutionary success-driven multigames},
\newblock \bibinfo{journal}{Europhysics Letters} \bibinfo{volume}{108}
  (\bibinfo{year}{2014}) \bibinfo{pages}{28004}.
%Type = Article
\bibitem[{Tu(2018)}]{tu2018contribution}
\bibinfo{author}{J.~Tu},
\newblock \bibinfo{title}{Contribution inequality in the spatial public goods
  game: Should the rich contribute more?},
\newblock \bibinfo{journal}{Physica A: Statistical Mechanics and its
  Applications} \bibinfo{volume}{496} (\bibinfo{year}{2018})
  \bibinfo{pages}{9--14}.
%Type = Article
\bibitem[{Nowak et~al.(2010)Nowak, Tarnita, and Antal}]{nowak2010evolutionary}
\bibinfo{author}{M.~A. Nowak}, \bibinfo{author}{C.~E. Tarnita},
  \bibinfo{author}{T.~Antal},
\newblock \bibinfo{title}{Evolutionary dynamics in structured populations},
\newblock \bibinfo{journal}{Philosophical Transactions of the Royal Society B:
  Biological Sciences} \bibinfo{volume}{365} (\bibinfo{year}{2010})
  \bibinfo{pages}{19--30}.
%Type = Article
\bibitem[{Su et~al.(2018)Su, Wang, and Stanley}]{su2018understanding}
\bibinfo{author}{Q.~Su}, \bibinfo{author}{L.~Wang}, \bibinfo{author}{H.~E.
  Stanley},
\newblock \bibinfo{title}{Understanding spatial public goods games on
  three-layer networks},
\newblock \bibinfo{journal}{New Journal of Physics} \bibinfo{volume}{20}
  (\bibinfo{year}{2018}) \bibinfo{pages}{103030}.
%Type = Article
\bibitem[{Su et~al.(2019)Su, Li, Wang, and Eugene~Stanley}]{su2019spatial}
\bibinfo{author}{Q.~Su}, \bibinfo{author}{A.~Li}, \bibinfo{author}{L.~Wang},
  \bibinfo{author}{H.~Eugene~Stanley},
\newblock \bibinfo{title}{Spatial reciprocity in the evolution of cooperation},
\newblock \bibinfo{journal}{Proceedings of the Royal Society B}
  \bibinfo{volume}{286} (\bibinfo{year}{2019}) \bibinfo{pages}{20190041}.
%Type = Article
\bibitem[{Wang and Szolnoki(2022)}]{wang2022reversed}
\bibinfo{author}{C.~Wang}, \bibinfo{author}{A.~Szolnoki},
\newblock \bibinfo{title}{A reversed form of public goods game: equivalence and
  difference},
\newblock \bibinfo{journal}{New Journal of Physics} \bibinfo{volume}{24}
  (\bibinfo{year}{2022}) \bibinfo{pages}{123030}.
%Type = Article
\bibitem[{Cheikbossian(2012)}]{cheikbossian2012collective}
\bibinfo{author}{G.~Cheikbossian},
\newblock \bibinfo{title}{The collective action problem: Within-group
  cooperation and between-group competition in a repeated rent-seeking game},
\newblock \bibinfo{journal}{Games and Economic Behavior} \bibinfo{volume}{74}
  (\bibinfo{year}{2012}) \bibinfo{pages}{68--82}.
%Type = Article
\bibitem[{Eckel et~al.(2016)Eckel, Fatas, Godoy, and Wilson}]{eckel2016group}
\bibinfo{author}{C.~C. Eckel}, \bibinfo{author}{E.~Fatas},
  \bibinfo{author}{S.~Godoy}, \bibinfo{author}{R.~K. Wilson},
\newblock \bibinfo{title}{Group-level selection increases cooperation in the
  public goods game},
\newblock \bibinfo{journal}{PloS One} \bibinfo{volume}{11}
  (\bibinfo{year}{2016}) \bibinfo{pages}{e0157840}.
%Type = Article
\bibitem[{Wang et~al.(2021)Wang, Xu, Chen, Yu, and He}]{wang2021inter}
\bibinfo{author}{J.~Wang}, \bibinfo{author}{W.~Xu}, \bibinfo{author}{W.~Chen},
  \bibinfo{author}{F.~Yu}, \bibinfo{author}{J.~He},
\newblock \bibinfo{title}{Inter-group selection of strategy promotes
  cooperation in public goods game},
\newblock \bibinfo{journal}{Physica A: Statistical Mechanics and its
  Applications} \bibinfo{volume}{583} (\bibinfo{year}{2021})
  \bibinfo{pages}{126292}.
%Type = Article
\bibitem[{Su et~al.(2016)Su, Li, Zhou, and Wang}]{su2016interactive}
\bibinfo{author}{Q.~Su}, \bibinfo{author}{A.~Li}, \bibinfo{author}{L.~Zhou},
  \bibinfo{author}{L.~Wang},
\newblock \bibinfo{title}{Interactive diversity promotes the evolution of
  cooperation in structured populations},
\newblock \bibinfo{journal}{New Journal of Physics} \bibinfo{volume}{18}
  (\bibinfo{year}{2016}) \bibinfo{pages}{103007}.
%Type = Article
\bibitem[{Su et~al.(2017)Su, Li, and Wang}]{su2017evolutionary}
\bibinfo{author}{Q.~Su}, \bibinfo{author}{A.~Li}, \bibinfo{author}{L.~Wang},
\newblock \bibinfo{title}{Evolutionary dynamics under interactive diversity},
\newblock \bibinfo{journal}{New Journal of Physics} \bibinfo{volume}{19}
  (\bibinfo{year}{2017}) \bibinfo{pages}{103023}.
%Type = Article
\bibitem[{Wang et~al.(2021)Wang, Pan, Ju, and He}]{wang2021public}
\bibinfo{author}{C.~Wang}, \bibinfo{author}{Q.~Pan}, \bibinfo{author}{X.~Ju},
  \bibinfo{author}{M.~He},
\newblock \bibinfo{title}{Public goods game with the interdependence of
  different cooperative strategies},
\newblock \bibinfo{journal}{Chaos, Solitons \& Fractals} \bibinfo{volume}{146}
  (\bibinfo{year}{2021}) \bibinfo{pages}{110871}.
%Type = Article
\bibitem[{Wang and Huang(2022)}]{wang2022between}
\bibinfo{author}{C.~Wang}, \bibinfo{author}{C.~Huang},
\newblock \bibinfo{title}{Between local and global strategy updating in public
  goods game},
\newblock \bibinfo{journal}{Physica A: Statistical Mechanics and its
  Applications} \bibinfo{volume}{606} (\bibinfo{year}{2022})
  \bibinfo{pages}{128097}.
%Type = Article
\bibitem[{Liu et~al.(2010)Liu, Li, Chen, and Wang}]{liu2010memory}
\bibinfo{author}{Y.~Liu}, \bibinfo{author}{Z.~Li}, \bibinfo{author}{X.~Chen},
  \bibinfo{author}{L.~Wang},
\newblock \bibinfo{title}{Memory-based prisoner’s dilemma on square
  lattices},
\newblock \bibinfo{journal}{Physica A: Statistical Mechanics and its
  Applications} \bibinfo{volume}{389} (\bibinfo{year}{2010})
  \bibinfo{pages}{2390--2396}.
%Type = Article
\bibitem[{Dong et~al.(2019)Dong, Xu, and Fan}]{dong2019memory}
\bibinfo{author}{Y.~Dong}, \bibinfo{author}{H.~Xu}, \bibinfo{author}{S.~Fan},
\newblock \bibinfo{title}{Memory-based stag hunt game on regular lattices},
\newblock \bibinfo{journal}{Physica A: Statistical Mechanics and its
  Applications} \bibinfo{volume}{519} (\bibinfo{year}{2019})
  \bibinfo{pages}{247--255}.
%Type = Article
\bibitem[{Shu et~al.(2018)Shu, Liu, Fang, and Chen}]{shu2018memory}
\bibinfo{author}{F.~Shu}, \bibinfo{author}{X.~Liu}, \bibinfo{author}{K.~Fang},
  \bibinfo{author}{H.~Chen},
\newblock \bibinfo{title}{Memory-based snowdrift game on a square lattice},
\newblock \bibinfo{journal}{Physica A: Statistical Mechanics and its
  Applications} \bibinfo{volume}{496} (\bibinfo{year}{2018})
  \bibinfo{pages}{15--26}.

\end{thebibliography}

\end{document}